\magnification=\magstep1
\baselineskip=20 true pt
\hsize=160 true mm
\vsize=200 true mm

\def\cen{\centerline}
\def\scri{{\cal I}}
\def\endproof{\nobreak\kern5pt\nobreak\vrule height4pt width4pt depth0pt}

\def\d{\partial}
\def\S{\Sigma}

\noindent
{\cen {\bf Topological Censorship and Higher Genus Black Holes}}
\bigskip
\cen{
{\it G.J.~Galloway},\footnote{$^1$}
{Dept. of Mathematics and Computer Science, University of Miami,
Coral Gables, FL 33124, USA. e-mail:galloway@math.miami.edu}
{\it K. Schleich},\footnote{$^2$}
{Dept. of Physics and Astronomy, University of British Columbia,
6224 Agriculture Road, Vancouver, BC, Canada V6T 1Z1. e-mail:
schleich@noether.physics.ubc.ca}
{\it D.M.~Witt},\footnote{$^3$}
{Dept. of Physics and Astronomy, University of British Columbia,
6224 Agriculture Road, Vancouver, BC, Canada V6T 1Z1. e-mail:
donwitt@noether.physics.ubc.ca}
and {\it E. Woolgar}\footnote{$^4$}
{Dept. of Mathematical Sciences and Theoretical Physics Institute,
University of Alberta, Edmonton, AB, Canada  T6G 2G1. e-mail:
ewoolgar@math.ualberta.ca}
}

\bigskip
{\cen {\bf Abstract}}
\smallskip
\noindent
Motivated by recent interest in black holes whose asymptotic geometry
approaches that of anti-de Sitter spacetime, we give a  proof of 
topological censorship applicable to spacetimes with such asymptotic 
behavior. Employing a useful rephrasing of topological censorship 
as a property of homotopies of arbitrary loops, we then explore the 
consequences of topological censorship for horizon topology of black 
holes. We find that the genera of horizons are controlled by the genus 
of the space at infinity. Our results make it clear 
that there is no conflict between topological censorship and the 
non-spherical horizon topologies of locally anti-de Sitter black holes. 

More specifically, let ${\cal D}$ be the domain of outer communications
of a boundary at infinity ``scri.'' We show that the Principle of 
Topological Censorship (PTC), that every causal curve in ${\cal D}$ having 
endpoints on scri can be deformed to scri, holds under reasonable 
conditions for timelike scri, as it is known to do for a simply
connected null scri. We then show that the PTC implies that the fundamental 
group of scri maps, via inclusion, onto the fundamental group of ${\cal D}$, 
{\it i.e.}, {\it every\/} loop in ${\cal D}$ is homotopic to a loop in scri. 
We use this to determine the integral homology of preferred spacelike 
hypersurfaces (Cauchy surfaces or analogues thereof) in the domain of 
outer communications of any 4-dimensional spacetime obeying the PTC. From 
this, we establish that the sum of the genera of the cross-sections in 
which such a hypersurface meets black hole horizons is bounded above by 
the genus of the cut of infinity defined by the hypersurface. Our results 
generalize familiar theorems valid for asymptotically flat spacetimes 
requiring simple connectivity of the domain of outer communications and 
spherical topology for stationary and  evolving black holes.
\par\vfil\eject

\noindent
{\cen {\bf Topological Censorship and Higher Genus Black Holes}}
\bigskip
\noindent
{\bf I. Introduction}
\smallskip
\noindent
It is generally a matter of course that the gross features of shape,
{\it i.e.}, the topology, of composite objects, from molecules to
stars, are determined by their internal structure. Yet black holes are
an exception. A black hole has little internal structure, but the
topology of its horizon is nonetheless strongly constrained, seemingly
by the {\it external} structure of spacetime. This was made apparent 
in early work of Hawking [1], who established via a beautiful variational 
argument the spherical topology of stationary horizons.  Hawking's proof 
was predicated on the global  causal theoretic result that no outer 
trapped surfaces can exist outside the black hole region, unless energy 
conditions or cosmic censorship are violated.  As argued by Hawking in 
[1], the possibility of toroidal topology, which arises as a borderline 
case in his argument, can be eliminated by consideration of a certain 
characteristic initial value problem and the assumption of analyticity; 
see [2] for further discussion of this issue.

In recent years an entirely different approach to the study of black
hole topology has developed, based on the notion of {\it topological
censorship}. In 1994, Chru\'sciel and Wald [3], improving in the
stationary setting the result on black hole topology considered in [2]
and [4], were able to remove the analyticity assumption in Hawking's
theorem by making use of the active topological censorship theorem of
Friedman, Schleich, and Witt [5] (hereinafter, FSW). This latter result
states that in a globally hyperbolic, asymptotically flat (hereinafter,
AF) spacetime obeying an Averaged Null Energy Condition (ultimately, in 
the modified form used below---see [6]), every causal curve beginning 
and ending on the boundary-at-infinity could be homotopically deformed 
to that boundary. When topological censorship holds, the domain of outer
communications (or DOC---the region exterior to black and white holes) of an AF 
spacetime must be simply connected [6]. 

Jacobson and Venkataramani [7], also using the topological censorship
theorem of FSW, were able to extend the result of Chru\'sciel and Wald on
black hole topology beyond the stationary case. The principle behind their
arguments  was that any horizon topology other than spherical would allow
certain causal curves outside the horizon to link with it, and so such
curves would not be deformable to infinity, which would contradict FSW.

In the early 1990s, new solutions with non-spherical black hole horizons 
were discovered in locally anti-de Sitter (adS) spacetimes [8--13]; for 
a recent review, see [14]. 
The original topological censorship theorem 
did not apply to these spacetimes since they were not AF and not
globally hyperbolic. However, two improvements to the proof of
topological censorship indicated that these differences ought not
matter. Galloway [15] was able to produce a ``finite infinity''
version of topological censorship that replaced the usual asymptotic
conditions on the geometry with a mild geometrical condition
on a finitely-distant boundary, and Galloway and Woolgar [16] were
able to replace the assumption of global hyperbolicity by weak cosmic
censorship. Moreover, it was soon observed [17] that topological
censorship in the sense of FSW held true for
each of the newly discovered black hole constructions in locally adS
spacetime, although
no general proof was known in this setting and
obviously the aforementioned corollary implying
spherical horizon topology could not hold. This is not paradoxical---the
topology of the locally adS black hole spacetimes is such that no
causal curve links with a non-spherical
horizon in such a way as to preclude a homotopy
deforming that curve to infinity.\footnote{$^5$}
{That topological censorship proofs should generalize to spacetimes
having the asymptotic structure of these new black holes was first
suggested in [7].}
In fact, we will see below that the corollary implying spherical
topology of black holes in  AF spacetime is merely a
special case
of a more general corollary of topological censorship which gives
a relationship between the topology of the black hole horizons
and the ``topology of
infinity.'' In this sense, the topology of the black hole horizons
 is governed by a
structure that is as ``external'' as possible, being entirely at
infinity.

In this paper, we will consider spacetimes that obey the Principle of
Topological Censorship (PTC). Let $\cal M$ be a spacetime  with metric 
$g_{ab}$. Suppose this spacetime can be conformally included into a 
spacetime-with-boundary $\cal M'= \cal M \cup \scri$, with metric $g_{ab}'$ 
whose restriction to ${\cal M}$ obeys $g_{ab}'=\Omega^2 g_{ab}$, and where 
$\scri$ is the $\Omega=0$ surface, which defines $\scri$ as the 
boundary-at-infinity. For a connected boundary component ${\scri}$,
\footnote{$^6$}
{For the purpose of discussion, in the AF case we assume for
simplicity
that $i^0$ is included in $\scri$.}
the domain of outer communications ${\cal D}$ is defined by
$${\cal D}:=I^+({\scri})\cap I^-({\scri})\quad .\eqno{(1)}$$
The PTC is the following condition on ${\cal D}$:
\smallskip
\noindent
{\bf Principle of Topological Censorship (PTC).} {\sl Every causal curve 
whose initial and final endpoints belong to ${\scri}$ is fixed endpoint
homotopic to a curve on ${\scri}$.}\footnote{$^7$}
{FSW's original statement of topological censorship required every causal
curve beginning and ending on ${\scri}$ to be homotopic to one in a
``simply connected neighborhood of infinity.'' Our phrasing above
accomodates spacetimes for which ${\scri}$ has no simply connected
neighborhoods.}

\smallskip
The PTC has already been established for general, physically reasonable 
AF spacetimes, {\it cf.}~[5], [16].
In the next section we present a proof of the PTC in a setting that 
includes many asymptotically locally anti-de Sitter black hole
spacetimes. 
This generalization exploits the fact that PTC proofs generally follow
from a condition on double-null components of the Ricci tensor. These 
components can be related to the double-null components of the stress 
energy tensor through the double-null components of the Einstein equations. 
This relation involves no trace terms and so clearly is insensitive to 
the cosmological constant.
This corrects the impression that the PTC is invalid in the presence 
of a negative cosmological constant [13,18],
an impression that, if it were true, would imply that the PTC 
in this case would impose no constraints at all on the topology of 
black hole horizons. 
Hawking's $g=0$ restriction [1,19,20] on horizon topology, by 
way of contrast, does assume non-negative energy density on the 
horizon and so is not in force in the presence of a negative
cosmological constant in vacuum but, as we will see, the PTC remains 
valid, and imposes constraints on the topology.

In Section III, we will prove that the mapping from the fundamental group 
of $\scri$ to that of the domain of outer communications is a surjection 
for spacetimes satisfying the PTC. This theorem generalizes previous results 
on the DOC being simply connected in asymptotically flat spacetimes. This 
characterization allows us to shift attention from causal curves to arbitrary 
loops in further study of the consequences of the PTC. 

In Section IV, by
considering loops restricted to certain spacelike surfaces and using
arguments from algebraic topology, we give a very direct derivation of
the topology of black hole horizons and the integral homology of 
hypersurfaces in exterior regions. More precisely, we consider the
topology of the closure of the Cauchy surfaces or analogues thereof
for the DOC whose intersections with the 
horizons are closed 2-manifolds
(``good  cuts'').
We find that
$$\sum_{i=1}^k g_i\le g_0$$
where the $g_i$ are the genera of the
cuts of the 
black hole horizons
and $g_0$ is the genus of a cut of $\scri$ by the surface. Thus the 
topology of the black hole horizons is constrained by the topology at infinity. 
This result also pertains to any sub-domain of the DOC that lies to the 
future of a cut of $\scri$ and  
whose Cauchy surfaces---or analogues thereof---meet the horizons
at closed 2-manifolds. Therefore, it applies to the topology of the black hole 
horizons in the presence of  black hole formation or evolution at times 
for which the appropriate sub-domain and surfaces can be found.
Finally we demonstrate that the integral homology of these 
surfaces is torsion free
and consequently completely determined by the Betti numbers. Furthermore,
these are completely fixed in terms of the genera of the boundary.

Section V contains a discussion concerning these results and their 
applicability to the case of non-stationary black holes.

\vfill\eject
\noindent
{\bf II. Validity of the PTC}
\smallskip
\noindent
The aim of this section is to present a version of the PTC applicable to
spacetimes which are asymptotically locally anti-de Sitter. Hence, we 
consider a spacetime $\cal M$, with metric $g_{ab}$, which can be
conformally included into a spacetime-with-boundary
$\cal M'= \cal M \cup \scri$, with metric $g_{ab}'$, such that $\partial 
\cal M' = \scri$ is timelike ({\it i.e.}, is a Lorentzian hypersurface 
in the induced metric) and $\cal M = \cal M' \setminus \scri$. 
We permit $\scri$ to have multiple components. With regard to the conformal
factor $\Omega\in C^{1}(\cal M')$, we make the standard assumptions that

\item{(a)} {$\Omega > 0$ and $g_{ab}' = \Omega^2 g_{ab}$  on $\cal M$, and}

\item{(b)} {$\Omega = 0$ and $d \Omega \ne 0$ pointwise on $\scri$.}

Just as in the case of spacetimes without boundary, we say that a
spacetime-with-boundary $\cal M'$ is {\it globally hyperbolic} provided 
$\cal M'$ is strongly causal and the sets $J^+(p,{\cal M'})\cap 
J^-(q,\cal M')$ are compact for all $p,q\in \cal M'$. Note that 
when $\scri$ is timelike, as in the situation considered here,
${\cal M}$ can never be globally hyperbolic.  However in many 
examples of interest, such as anti-de Sitter space and the domain 
of outer communications of various locally adS spaces, $\cal M'$ is.

For later convenience, we  define a Cauchy surface for $\cal M'$ to 
be a subset $V'\subset \cal M'$ which is met once and only once by each 
inextendible causal curve in $\cal M'$. Then $V'$ will be a spacelike 
hypersurface which, as a manifold-with-boundary, has boundary on $\scri$. It 
can be shown, as in the standard case, that a spacetime-with-timelike-boundary
$\cal D'$ which is globally hyperbolic admits a Cauchy surface $V'$ and is 
homeomorphic to $R\times V'$. (This can be shown by directly modifying the 
proof of Prop.~6.6.8 in [19].  Alternatively, arguments can be given which 
involve invoking Prop.~6.6.8.  For example, one can  consider the ``doubled 
spacetime'' $\cal M''$ of $\cal M'$ through $\scri$, with metric on this 
double defined by the natural extension of that on $\cal M'$. Then one can apply Prop.~6.6.8 
to $\cal M''$, which will be globally hyperbolic if $\cal M'$ is.) Many of 
the locally anti-de Sitter and related models [8--11,13,14] which have been 
constructed have DOCs which admit Cauchy surfaces of this sort.

The proof of the PTC is a consequence of the following basic result.

\smallskip
\noindent
{\bf Theorem 2.1.} {\sl Let ${\cal M}\subset \cal M'$ be as described
above, and assume the
following conditions hold.

\item{({\it i})}{$\cal M'= M \cup \scri$ is globally hyperbolic.}
\item{({\it ii})}{There is a component $\scri_0$ of $\scri$ which 
admits a compact spacelike cut.}
\item{({\it iii})}{For each point $p$ in $\cal M$ near ${\scri}_0$ 
and any future complete null geodesic $s\to\eta(s)$ in $\cal M$ 
starting at $p$, $\int_0^{\infty}{\rm Ric}(\eta',\eta')\,ds\ge 0$.}

\noindent
Then $\scri_0$ cannot communicate with any other component of $\scri$,
{\it i.e.}, $J^+(\scri_0)\cap
(\scri\setminus \scri_0) = \emptyset$.
}
\smallskip

Condition {({\it iii})} is a modified form of the Average
Null Energy Condition (ANEC). This term
usually refers to a condition of the same form as ({\it iii}) except that the 
integral is taken over geodesics complete to both past and future.
Note that if one assumes that the Einstein equations
with cosmological constant
$R_{ab} - {1\over 2} R g_{ab} + \Lambda g_{ab} = 8\pi T_{ab}$
hold, then for any null vector X, ${\rm Ric}(X,X) = R_{ab} X^aX^b 
= 8\pi T_{ab}X^aX^b$. Then the integrand ${\rm Ric}(\eta',\eta')$ 
in ({\it iii}) could be replaced by $ T(\eta',\eta')$. 
Clearly, the presence and sign of the 
cosmological constant is irrelevant to whether or not a spacetime 
satisfying the Einstein equations will satisfy condition ({\it iii}). 

For the next theorem, if $\d {\cal M}'$ is not connected, let $\scri$ 
denote a single component of $\d {\cal M}'$.  Let ${\cal D}=I^+({\scri})
\cap I^-({\scri})$ be the domain of outer communications of $\cal M$ 
with respect to $\scri$. Assume that $\cal D$ does not meet any other 
components of $\d \cal M'$. Note that if the conditions of Theorem 2.1 
hold, this latter assumption is automatically satisfied. Using in
addition the fact that $\scri$ is timelike, it follows that
$\cal D$ is connected and the closure of $\cal D$ in $\cal M'$
contains $\scri$.   Then $\cal D'\cal := {\cal D} \cup \scri$ is a 
connected spacetime-with-boundary, with $\partial \cal D' = \scri$ 
and $\cal D = \cal D' \setminus \scri$.

We now state the following topological censorship theorem, applicable to
asymptotically locally adS spacetimes.

\smallskip
\noindent {\bf Theorem 2.2.} {\sl Let $\cal D$ be the domain of outer
communications with respect to
$\scri$ as described above,  and assume the following conditions hold.

\item{({\it i})}{$\cal D'= D\cup \scri$ is globally hyperbolic.}
\item{({\it ii})}{$\scri$ admits a compact spacelike cut.}
\item{({\it iii})}{For each point $p$ in $\cal M$ near $\scri$ and any
future complete null
geodesic $s\to\eta(s)$ in $\cal D$ starting at $p$, $\int_0^{\infty}{\rm
Ric}(\eta',\eta')\,ds\ge
0$.}

\noindent Then the PTC holds on ${\cal D}$. }

\smallskip
\noindent 
{\bf Remark.} Let $K$ be a cut of $\scri$, and let $\scri_K$ be
the portion of $\scri$
to the future of $K$, $\scri_K = \scri \cap I^+(K)$.  Let  $\cal D_K$ be
the domain of
outer of communications with respect to $\scri_K$,  ${\cal D_K} =
I^+(\scri_K)\cap I^-(\scri_K)
= I^+(K) \cap I^-(\scri)$.
Theorems 2.1 and 2.2 apply equally as well to $\cal D_K$.  This procedure,
first discussed by
Jacobson and Venkataramani [7], allows one, by the methods of this paper,
to study the topology
of cuts on the future event horizon ${\cal H} = \partial I^-(\scri)$ of the
form $\partial I^+({\scri_K})
\cap\cal H$.  By taking $K$ sufficiently far to the future, this procedure
enables one to consider cuts on
$\cal H$ well to the future of the initial formation of the black hole,
where one has a greater
expectation that the intersection of
$\cal H$ with $\partial I^+({\scri_K})$ will be reasonable ({\it i.e.}, 
a surface).  See [7], and
Sections 4 and 5 below for
further discussion of this point. In what follows, it is worth keeping in
mind that
$\cal{I}$ may refer to the
portion of scri to the future of a cut.

\smallskip\noindent
{\bf Proof of Theorem 2.1:} The global hyperbolicity of $\cal M'$ implies
that $\scri_0$ is strongly causal
as a spacetime in its own right, and that the sets $J^+(x,\scri_0)\cap
J^-(y,\scri_0)$
($\subset J^+(x,{\cal M'})\cap J^-(y, \cal M')$) 
for $x,y \in \scri_0$ have compact closure in
$\scri_0$.  This is sufficient
to imply that
$\scri_0$ is globally hyperbolic as a spacetime in its own right.
Assumption ({\it ii}) then implies that
$\scri_0$ is foliated by compact Cauchy surfaces.

Now suppose that $J^+(\scri_0)$ meets some other component $\scri_1$ of
$\scri$, {\it i.e.}, suppose there
exists a future directed causal curve from a point $p\in \scri_0$ to a
point $q\in \scri_1$.  Let
$\S_0$ be a Cauchy surface for $\scri_0$ passing through $p$. Push $\S_0$ 
slightly in the normal direction to $\scri_0$ to obtain a compact spacelike 
surface $\S$ contained in $\cal M$.  Let $V$ be a compact
spacelike hypersurface-with-boundary spanning $\S_0$ and $\S$.  By
properties of the conformal factor
we are assured, for suitable pushes, that $\S$ is {\it null mean
convex\/}.  By this we mean
that the future directed null geodesics issuing orthogonally from $\S$
which are ``inward pointing" with respect to $\scri_0$ ({\it i.e.}, 
which point away from $V$) have negative divergence.

Under the present supposition, $J^+(V)$ meets $\scri_1$.  At the same time,
$J^+(V)$ cannot contain
all of $\scri_1$. If it did, there would exist a past inextendible causal
curve in
$\scri_1$ starting at $q\in\scri_1$ contained in the compact set
$J^+(V)\cap J^-(q)$, contradicting
the strong causality of $\cal M'$. Hence, $\d J^+(V)$ meets $\scri_1$; let
$q_0$ be a point in $\d J^+(V)\cap\scri_1$. 
Since $\cal M'$ is globally hyperbolic, it is causally simple; 
{\it cf.}~Proposition 6.6.1 in [19], which remains valid in the 
present setting. Hence, $\partial J^+(V) = J^+(V) \setminus I^+(V)$, 
which implies that there exists a future directed null curve $\eta\subset
\d J^+(V)$ that extends from a point on $V$ to $q_0$. (Alternatively,
one can prove the existence of this curve from results of [21], which 
are also valid in the present setting.)
It is possible that $\eta$ meets $\scri$ several times
before reaching $q_0$.
Consider only the portion $\eta_0$ of $\eta$ which extends from the initial
point of $\eta$ on $V$ up to,
but not including, the first point at which $\eta$ meets $\scri$.

By properties of achronal boundaries and the conformal factor, $\eta_0$ is
a future complete
null geodesic in $\cal M$ emanating from a point on $V$.  Since $\eta_0$
cannot enter $I^+(V)$,
it follows that: (1) $\eta_0$ actually meets $V$ at a point $p_0$ of $\S$,
(2) $\eta_0$ meets
$\S$ orthogonally at $p_0$, and (3) $\eta_0$ is inward pointing with
respect to $\scri_0$
({\it i.e.}, $\eta_0$ points away from $V$).  Since $\eta_0$ is future 
complete, the energy condition
({\it iii}) and null mean convexity of $\S$ imply that there is a null focal
point to $\S$ along $\eta_0$.
But beyond the focal point, $\eta_0$ must enter $I^+(\S)$ ($\subset
I^+(V)$), contradicting
$\eta_0 \subset \d J^+(V)$. \hfill\endproof
\smallskip

\smallskip\noindent
{\bf Proof of Theorem 2.2:} The proof is an application of
Theorem 2.1, together with a covering space argument. Fix $p\in\scri$.
The inclusion map
$i:\scri\to\cal D'$ induces a homomorphism of fundamental groups
$i_*:\Pi_1(\scri,p)\to
\Pi_1({\cal D'},p)$.  The image $G =i_*(\Pi_1(\scri,p))$ is a subgroup of
$\Pi_1({\cal D'},p)$.  Basic covering space theory guarantees that there
exists an
essentially unique covering space $\widetilde{\cal D}'$ of $\cal D'$ such that
$\pi_*(\Pi_1(\widetilde{\cal D}',\widetilde{p}))= G =i_*(\Pi_1(\scri,p))$,
where $\pi:\widetilde{\cal D}'\to \cal D'$ is the covering map, and
$\pi(\widetilde{p})
=p$.  Equip $\widetilde{\cal D}'$ with the pullback metric $\pi^*(g')$ so that
$\pi:\widetilde{\cal D}'\to \cal D'$ is a local isometry.  We note that when
$\scri$ is simply connected, $\widetilde{\cal D}'$ is the universal cover of
$\cal D'$, but in general it will not be.  In a somewhat different context
({\it i.e.}, when $\cal D'$ is a spacetime without boundary and $\scri$ is a
spacelike hypersurface), $\widetilde{\cal D}'$ is known as the Hawking
covering spacetime, {\it cf.}~[19], [22].

The covering spacetime  $\widetilde{\cal D}'$ has two basic properties:

\item{(a)}{The component $\widetilde{\scri}$ of $\pi^{-1}(\scri)$} which 
passes through $\widetilde{p}$ is a copy of $\scri$, {\it i.e.},
$\pi_*|_{\widetilde{\scri}}: \widetilde{\scri}\to \scri$ is an isometry.
\item{(b)}$i_*(\Pi_1(\widetilde{\scri},
\widetilde{p}))=\Pi_1(\widetilde{\cal D}',\widetilde{p})$, where $i:
\widetilde{\scri}
\hookrightarrow \widetilde{\cal D}'$.

Property (b) says that any loop in $\widetilde{\cal D}'$ based at
$\widetilde{p}$
can be deformed through loops based at  $\widetilde{p}$ to a loop in
$\widetilde{\scri}$ based at $\widetilde{p}$.  This property is an easy
consequence
of the defining property
$\pi_*(\Pi_1(\widetilde{\cal D}',\widetilde{p}))=i_*(\Pi_1(\scri,p))$ 
and the homotopy lifting property.  In turn, property (b) easily 
implies that any curve in $\widetilde{\cal D}'$ with endpoints on 
$\widetilde{\scri}$ is fixed endpoint homotopic
to a curve in $\widetilde{\scri}$.

Now let $\gamma$ be a future directed causal curve in $\cal D'$ with 
endpoints on $\scri$.
Assume $\gamma$ extends from $x\in\scri$ to $y\in\scri$.
Let $\widetilde{\gamma}$ be the lift of $\gamma$ into $\widetilde{\cal D}'$
starting at
$\widetilde x\in \widetilde{\scri}$.
Note that the hypotheses of Theorem 2.1, with ${\cal M}=\widetilde{\cal D}
:=\pi^{-1}(D)$, $\cal M'= \widetilde{\cal D}'$, and $\scri_0=\widetilde{\scri}$,
are satisfied.  Hence the future endpoint $\widetilde y$ of
$\widetilde{\gamma}$
must also lie on $\widetilde{\scri}$.  Then we know that $\widetilde{\gamma}$
is fixed endpoint homotopic to a curve in $\widetilde{\scri}$.  Projecting
this homotopy down to $\cal D'$, it follows that $\gamma$ is fixed endpoint
homotopic to a curve in $\scri$, as desired. \hfill\endproof
\smallskip

\smallskip
\noindent
{\bf Remarks.\/}
Since many examples of locally adS spacetimes, for example those 
constructed by identifications of adS [9--11,13,14], do not obey the generic 
condition, our aim was to present a version of the PTC which does not 
require it. However if one replaces the energy condition ({\it iii}) by 
the generic condition and the ANEC: $\int_{-\infty}^{\infty}{\rm Ric}
(\eta',\eta')\,ds\ge 0$ along any complete null geodesic $s\to\eta(s)$ 
in $\cal D$, then Theorems 2.1 and 2.2 still hold, and moreover they do 
not require the compactness condition ({\it ii}). The proof of Theorem 2.1 
under these new assumptions involves the construction of a complete null 
line (globally achronal null geodesic) which is incompatible with the 
energy conditions. The results in this setting are rather general, and 
in particular the proofs do not use in any essential way that $\scri$ is 
timelike. We also mention that the global hyperbolicity assumption in 
Theorem 2.2 can be weakened in a fashion similar to what was done in the 
AF case in [16].

\smallskip
We now turn
our attention to a characterization of topological censorship which
will allow us to more easily explore the consequences of topological
censorship for black hole topology.

\bigskip
\noindent
{\bf III. Algebraic Characterization of Topological Censorship}
\smallskip
\noindent
The main result of this section is a restatement of the properties of
a spacetime satisfying the PTC in a 
language amenable to algebraic topological considerations.
\noindent

Let ${\cal D}$ and ${\scri}$ be as in Section I. Then
$\cal D' = \cal D \cup\scri$ is a spacetime-with-boundary, 
with $\partial\cal D' = \scri$.
The inclusion map $i:\scri \to \cal D'$ induces a homomorphism
of fundamental groups  $i_*:\Pi_1({\scri})\to\Pi_1(\cal D')$. Then

\smallskip
\noindent
{\bf Proposition 3.1.} {\sl If the PTC holds for ${\cal D'}$, 
then the group homomorphism
$i_*:\Pi_1({\scri})\to\Pi_1({\cal D'})$ induced by inclusion is surjective.}
\smallskip
\noindent
{\bf Remark.} Note that the fundamental groups of $\cal D$ and
$\cal D'$ are trivially isomorphic, $\Pi_1(\cal D) \cong$ $\Pi_1(\cal D')$.
Hence,
Proposition 3.1 says roughly that every loop in $\cal D$ is deformable to
a loop in $\scri$.
Moreover, it implies that $\Pi_1({\cal D})$ is isomorphic
to  the factor group $\Pi_1({\scri})/{\rm ker}\,i_*$.  In particular, if
$\scri$ is simply connected
then so is $\cal D$.
\smallskip

Our proof relies on the following straightforward lemma in topology:
\smallskip
\noindent
{\bf Lemma 3.2.} {\sl Let $N$ be a manifold and $S$ an embedded
submanifold with inclusion mapping $i:S\to N$. If in the universal
covering space of $N$ the inverse image of $S$ by the covering map
is connected, then $i_*:\Pi_1(S)\to\Pi_1(N)$ is surjective.}

\smallskip
\noindent
{\bf Proof:}  Strictly speaking, we are dealing with pointed spaces $(S,p)$
and $(N,p)$,
for some fixed point $p\in S$, and we want to show that
$i_*:\Pi_1(S,p)\to\Pi_1(N,p)$ is onto.
Let $[c_0]$ be an element of  $\Pi_1(N,p)$, {\it i.e.}, let $c_0$ be a
loop in $N$ based at $p$.
Let $\tilde N$ be the universal covering space of $N$, with covering map
$\pi: \tilde N\to N$.
Choose $\tilde p\in \tilde N$ such that $\pi(\tilde p) = p$.
Let $\tilde c_0$ be the lift of $c_0$ starting at $\tilde p$; then $\tilde
c_0$ is a curve in  $\tilde N$
extending from $\tilde p$ to a point $\tilde q$ with $\pi(\tilde q)= p$.
Since $\tilde p,
\tilde q \in \pi^{-1}(S)$
and $\pi^{-1}(S)$ is path connected, there exists a curve $\tilde c_1$ in
$\pi^{-1}(S)$ from $\tilde p$
to $\tilde q$.  The projected curve $c_1 = \pi(\tilde c_1)$ is a loop in
$S$ based at $p$, {\it i.e.},
$[c_1] \in \Pi_1(S,p)$.  Since $\tilde N$ is simply connected, $\tilde c_0$
is fixed endpoint
homotopic to $\tilde c_1$.  But this implies that $c_0$ is homotopic to
$c_1$ through loops based
at $p$, {\it i.e.}, $i_*([c_1]) = [c_0]$, as desired.
\hfill\endproof

\smallskip
\noindent
{\bf Proof of Proposition 3.1:}
We let ${\widetilde N}$ be the universal
covering  spacetime of $\cal N:=\cal D'$ with projection
$\pi:{\widetilde {\cal N}}
\to{\cal N}$. 
Then ${\widetilde {\cal N}}=
{\widetilde {\cal D}}
\cup {\widetilde {\scri}}$, where
${\widetilde {\cal D}}=\pi^{-1}(\cal D)$ is the universal covering
spacetime of ${\cal D}$ and ${\widetilde {\scri}}= \pi^{-1}(\scri)$ is
the boundary
of ${\widetilde N}$.

Every point in ${\widetilde {\cal D}}$ belongs to the inverse
image by $\pi$ of some point in ${\cal D}$, and every point in ${\cal D}$
lies on some causal curve beginning and ending on $\scri$, so every point
in ${\widetilde {\cal D}}$ lies on some causal curve beginning and ending
on ${\widetilde {\scri}}$. By the PTC  and basic lifting properties,
no such curve can end on a different
connected component of ${\widetilde {\scri}}$ than it began on. Therefore,
if we label these connected components by $\alpha$, the open sets
$I^+({\widetilde {\scri_{\alpha}}})\cap I^-({\widetilde {\scri_{\alpha}}})$
are a disjoint open cover of ${\widetilde {\cal D}}$. But ${\widetilde
{\cal D}}$ is connected, so $\alpha$ can take only one value, whence
${\widetilde {\scri}}$ is connected. 
It follows that $\cal D'$ satisfies the conditions of Lemma 3.2 
with $S=\scri$ and therefore 
$i_*:\Pi_1({\scri})\to
\Pi_1({\cal D'})$ is surjective. Thus $i_*:\Pi_1({\scri})\to
\Pi_1({\cal D})$ is surjective.\hfill\endproof

\smallskip
All known locally anti-de Sitter black holes and related spacetimes are
in accord with Proposition 3.1.
We conclude this section with the following simple corollary to Proposition 3.1.

\smallskip
\noindent
{\bf Corollary 3.3.} {\sl If the PTC holds for ${\cal D'}$
then $\cal D$ is orientable if ${\scri}$ is.}

\smallskip
\noindent
{\bf Proof:}
In fact  $\cal D'$ is orientable, for
if $\cal D'$ were not orientable and $\scri$ were, then $\cal D'$
would possess an orientable double cover containing two copies ${\scri}_1$
and ${\scri}_2$
of $\scri$.  Then a curve from $p_1 \in {\scri}_1$ to $p_2 \in {\cal
I}_2$, where
$\pi(p_1)=\pi(p_2)=p\in\scri$, would project to a loop in $\cal D'$ not
deformable  to $\scri$, contrary to the surjectivity of $i_*$.
\hfill\endproof

\bigskip

\noindent
{\bf IV. Application to Black Hole Topology}
\smallskip
\noindent
The boundary of the region of spacetime visible to
observers at $\scri$ by future directed causal curves 
is referred to as the
event horizon.
This horizon is a set of one or more null surfaces, also called black hole horizons,
 generated by null geodesics
that have 
no future endpoints but possibly 
have past endpoints. The topology of 
these black hole horizons is constrained in spacetimes obeying the PTC 
because, as was seen in Section III,
the topology of the domain of outer
communications is constrained by the PTC---intuitively, causal curves 
that can communicate with observers at $\scri$ cannot link with these 
horizons in a non-trivial way, rather they only carry information
about the non-triviality of curves on $\scri$.

Useful though this 
description
is, it does not characterize the topology of these horizons themselves, 
but rather the hole that their excision leaves in 
the spacetime. However, 
one can obtain certain information about these horizons if one 
considers the topology of the intersection of certain spacelike
hypersurfaces with the horizons, those whose intersection with the horizons 
are closed spacelike 2-manifolds (good cuts of the horizons).
For example, if one has a single horizon of product form, as is the case for
a stationary black hole, then the horizon topology is determined by 
that of the 2-manifold. Of course,  black hole horizons are generally not of 
product form; however, the topology of any good cut is still closely 
related to that of the horizons. For example, if each horizon has a region 
with product topology, this topology is determined by that of a good 
cut passing through this region. As demonstrated by Jacobson and 
Venkataramani [7] for asymptotically flat spacetimes, the PTC constrains 
the topology of  good cuts of the horizons by the closure of a Cauchy 
surface for the domain of outer communications to be  2-spheres.

Now it is possible that a black hole horizon admits two good
cuts by spacelike hypersurfaces, one entirely to the future of the other,
such that the cuts are not 
homeomorphic. 
This can only happen if null
generators enter the horizon between the two cuts.
Physically, such a situation corresponds to a black
hole with transient behavior, such as that induced by formation from
collapse, collision of black holes
or absorption of matter.
Jacobson and Venkataramani observed that their theorem could also be 
applied to such situations for certain spacelike hypersurfaces that
cut the horizons sufficiently far away from regions of black hole formation,
collision or matter absorption. 
Precisely, when the domain of outer communications is globally
hyperbolic to the future of a cut of $\scri$ and if the PTC holds on this
sub-domain, then the PTC will constrain the topology of this sub-domain, 
of its Cauchy surface and ultimately of good cuts of the horizons.
Thus, though the PTC does not determine the topology of arbitrary
embedded hypersurfaces or the  cuts they make on the horizons, it does
do so for hypersurfaces 
homeomorphic to Cauchy surfaces for these sub-domains that make good cuts
of the horizons.
Such sub-domains can be found for black hole horizons that settle down at 
late times.

Below we provide a generalization of these results applicable to a more general
set of spacetimes satisfying the PTC than asymptotically flat 
spacetimes. 
This generalization is based on the observation that if one can 
continuously push any loop in the DOC down into an appropriate spacelike
surface that cuts the horizons in  spacelike 2-manifolds, it follows from
Prop.~3.1 that the fundamental group of the spacelike surface is related 
to that of $\scri$. 
In the interests of a clear presentation that carefully treats all technical
details, we first
prove the following results for globally hyperbolic 
spacetimes  with timelike $\scri$\footnote{$^8$}
{We remind the reader that the notions of global hyperbolicity and Cauchy
surfaces for spacetimes with timelike $\scri$ are reviewed at the beginning of Section
II.} and, we emphasize,
for any globally hyperbolic sub-domain corresponding to the
future of a cut of $\scri$. We then provide the results for the case
of globally hyperbolic, asymptotically flat spacetimes. From the method
of proof of these two cases,
it is manifestly apparent that these theorems can
be easily generalized to a 
wider class
of spacetimes that satisfy the PTC. We conclude with a remark about how
to provide the correct technical statement and proofs for these cases.
We will also comment further on the transient 
behavior associated with black hole
formation in the Discussion section.

Let $\cal M$ be a spacetime with timelike infinity $\scri$ and domain of outer
communications~$\cal{D}$.
Assume $\scri$ is connected and orientable.
Let $K$ be a spacelike cut of $\scri$, and let $\scri_K$ be the portion of
$\scri$
to the future of $K$, $\scri_K = \scri \cap I^+(K)$.  Let  $\cal D_K$ be
the DOC with respect to $\scri_K$,  ${\cal D_K} =
I^+(\scri_K)\cap I^-(\scri_K)
= I^+(K) \cap I^-(\scri)$.
Note ${\cal D}_K' := {\cal D}_K \cup \scri_K$ is a
connected spacetime-with-timelike-boundary, with $\partial \cal D_K' = \cal
I_K$
and $\cal D_K = \cal D_K' \setminus \scri_K$. 
In the following we will assume that 
$\cal D_K'$ is globally hyperbolic
and has a Cauchy surface $V'$ as in section II.
The following theorem is the main result pertaining to the topology of
black holes.

\smallskip
\noindent {\bf Theorem 4.1.} {\sl Let $\cal D_K$ be the domain of outer
communications to the future of the cut $K$ on $\scri$ as described above.
Assume
$\cal D_K'$ is globally hyperbolic 
and satisfies 
the PTC. Suppose $V'$ is a Cauchy surface for $\cal D_K'$ such that its
closure $V=\overline{V'}$ in $\cal M'$
is a compact topological $3$-manifold-with-boundary whose boundary
$\partial V$ (corresponding to the edge of $V'$ in $M'$) consists of
a disjoint union of compact 2-surfaces,
$$\partial V =\bigsqcup\limits_{i=0}^{k}\Sigma_i\quad ,\eqno{(2)} $$
where $\Sigma_0$ is on  $\scri$ and the $\Sigma_i$,
$i=1,\dots,k$, are on the event horizon.
Then
$$\sum_{i=1}^k g_i\le g_0 \quad ,\eqno{(3)}$$
where $g_j=$ the genus of $\Sigma_j$, $j=0,1,\dots,k$.
In particular, if $\Sigma_0$ is a $2$-sphere then so is each $\Sigma_i$, 
$i=1,\dots,k$.}

\smallskip
\noindent
{\bf Remark 1.\/}
Many known examples of
locally adS black hole spacetimes have black hole horizons with genus equal
of that of scri [9--11,13].
In fact, for these examples, $V$ is a product space.

\smallskip
\noindent
{\bf Remark 2.\/}
Theorem 4.1 has been stated for spacetimes with timelike
$\scri$. An analogous version for asymptotically flat spacetimes holds as
well, but
differs slightly in technical details.  The AF case will be considered later
in the  section.

\smallskip
Theorem 4.1 is established in a series of lemmas. Lemma 4.2 connects the 
fundamental group of the Cauchy surface to that of $\scri$. It is the 
only lemma that uses the conditions that the 
spacetime
is globally 
hyperbolic and satisfies the PTC. The remaining lemmas are based purely 
on algebraic topology and are in fact applicable to any 3-manifold with 
compact boundary.

\smallskip
\noindent
{\bf Lemma 4.2.} {\sl Let the setting be as in Theorem 4.1. Then the group
homomorphism
$i_*:\Pi_1(\Sigma_0)\to\Pi_1(V)$ induced by inclusion
$i:\Sigma_0\to V$ is onto.}

\smallskip
\noindent
{\bf Proof:}  Let $Z^a$ be a timelike vector field on $\cal D_K'$ tangent to
$\scri_K$.
Let $r: {\cal D_K'} \to V'$ be the continuous projection map sending each
point $p \in
\cal D_K'$ to the unique point in $V'$ determined by the integral curve of
$Z^a$ through
$p$.  Note that $r(\scri_K) = \Sigma_0$.

Fix $p \in \Sigma_0$.  All loops considered are based at $p$.
Let $c_0$ be a loop in $V$.  By deforming $c_0$ slightly we can assume
$c_0$ is in $V'$.  Since $\cal D_K'$ satisifies the hypotheses of Theorem 2.2,
the PTC holds for $\cal D_K'$. Then, by Proposition 3.1, $c_0$ can be
continuously
deformed through loops in
$\cal D_K'$ to a loop $c_1$ in $\scri_K$. It follows by composition with
$r$  that $c_0
= r\circ c_0$
can be continuously deformed through loops in V to the loop $c_2 =
r\circ c_1$ in $\Sigma_0$, and hence $i_*([c_2]) = [c_0]$, as
desired.\hfill\endproof
\smallskip

By Corollary 3.3 and the assumptions of Theorem 4.1,  $V$ is a
connected, orientable, compact 3-manifold-with-boundary whose
boundary consists of $k+1$ compact surfaces $\Sigma_i$, $i=0,..., k$.
For the following results, which are
purely topological, we assume $V$ is any such manifold.

All homology and cohomology below
is taken over the integers. The $j^{\rm th}$ homology group of a
manifold ${\cal P}$ will be denoted by $H_j({\cal P})$,
$b_j({\cal P}):=$ the rank of the free part of $ H_j({\cal P})$ will denote
the $j^{\rm th}$
Betti number of ${\cal P}$, and $\chi({\cal P})$ will denote the
Euler characteristic (the alternating sum of the Betti numbers).
In the case of $b_j(V)$ ($=b_j(V_0)$), we will simply write $b_j$.

In general, for $V$ as assumed above, $b_0=1$, $b_3=0$, and $b_1$ and $b_2$
satisfy the following inequalities:

\smallskip
\noindent
{\bf Lemma 4.3.} {\sl For $V$ as above, then
\item{(a)}{$\displaystyle b_1 \ge \sum_{i=0}^k g_i$, and}
\item{(b)}{$b_2 \ge k$.}
}

\smallskip
\noindent
{\bf Proof:}
The boundary surfaces $\Sigma_1$, $\Sigma_2$, $\dots$,
$\Sigma_k$ clearly determine $k$ linearly independent $2$-cycles in $V$,
and hence (b) holds. To prove (a) we use the formula, $\chi(V)={1\over 2}
\chi(\partial V)$, valid for any compact, orientable, odd-dimensional
manifold. This formula, together with the expressions,
$\chi(V) = 1-b_1+b_2$ and
$\chi(\partial V)=\sum\limits_{i=0}^k\chi(\Sigma_i) =  2(k+1) -2\sum
\limits_{i=0}^k g_i$, implies the equation
$$b_1 = b_2 + \sum_{i=0}^kg_i-k\quad . \eqno{(4)}$$
The inequality (a) now follows immediately from (b).\hfill\endproof

\smallskip
\noindent
{\bf Lemma 4.4.} {\sl If the group homomorphism
$i_*:\Pi_1(\Sigma_0)\to\Pi_1(V)$ induced by inclusion
is onto then the inequality (3), $\sum_{i=1}^k g_i\le g_0$, holds.
In particular, if $\Sigma_0$ is a $2$-sphere then so is each $\Sigma_i$, $i
= 1,... ,k$.}

\smallskip
\noindent
{\bf Proof:}
We use the fact that the first integral homology group of a space
is isomorphic to the fundamental group modded out by its
commutator subgroup.  Hence, modding out
by the commutator subgroups of $\Pi_1(\Sigma_0)$
and $\Pi_1(V)$, respectively, induces from $i_*$ a
surjective homomorphism from $H_1(\Sigma_0)$ to $H_1(V)$.
It follows that the rank of the free part of $H_1(V)$
cannot be greater than that of $H_1(\Sigma_0)$, {\it i.e.},
$$b_1\le b_1(\Sigma_0) = 2g_0\quad .\eqno{(5)}$$
Combining this inequality with the inequality (a) in Lemma 
4.3 yields the inequality (3). Since $V$ is orientable, so are its 
boundary components. If $\Sigma_0$ is a $2$-sphere, then each $\Sigma_i$, 
$i=1,\dots,k$, is forced by (3) to have genus zero, and hence is a
$2$-sphere.\hfill\endproof
\smallskip

\smallskip
\noindent
{\bf Proof of Theorem 4.1:} Follows immediately from Lemma 4.2
and Lemma~4.4.\hfill\endproof
\smallskip

Next we, show that the condition on $i_*$ in Lemma 4.4  completely
determines the homology of $V$.

\smallskip
\noindent
{\bf Proposition 4.5.} {\sl If $i_*:\Pi_1(\Sigma_0)\to \Pi_1(V)$ is onto,
then the integral homology $H_*(V,Z)$ is torsion free, and hence is completely
determined by the Betti numbers.  Furthermore, the inequalities in Lemma 4.3
become equalities,
\item{(a)}{$\displaystyle b_1 = \sum_{i=0}^k g_i$, and}
\item{(b)}{$b_2 = k$.}
}

\smallskip
\noindent
{\bf Proof:}  We first prove that $H_*(V,Z)$ is torsion free.  Since $V$ has
boundary, $H_3(V)=0$. Also
$H_0(V)$ is one copy of $Z$ as $V$ is connected.  Thus, we need to show
$H_1(V)$ and $H_2(V)$ are free.

\smallskip
\noindent
{\bf Claim 1.} $H_2(V)$ is free.

\smallskip
\noindent
To prove Claim 1 we recall the classic result that $H_{n-1}(N^n)$ is free
for an orientable
closed n-manifold $N^n$.  To make use of this, let $V'$ be a compact
orientable $3$-manifold
without boundary containing $V$ ({\it e.g.},
take $V'$ to be the double of $V$), and  let $B=V'\setminus V$.

Assume that $W$ is a non-trivial torsion element in $H_2(V)$.
Now view $W$ as an element in  $H_2(V')$.  Suppose $W=0$ in $H_2(V')$.
Then $W=0$ in $H_2(V',B)$. By
excision,
$H_2(V',B)=H_2(V,\partial V)$, where $\partial V$ is the manifold boundary
of $V$. Hence $W=0$ in
$H_2(V,\partial V)$. This means that $W=$ a sum of boundary components in
$H_2(V)$. But a sum of boundary components cannot be a torsion element.
Thus, $W\ne 0$ in $H_2(V')$. Moreover, if $nW = 0$ in $H_2(V)$ then
$nW = 0$ in $H_2(V')$. It follows that $W$ is a non-trivial torsion element
in $H_2(V')$, a contradiction. Hence, $H_2(V)$ is free.

\smallskip
\noindent
{\bf Claim 2.} $H_1(V)$ is free.

\smallskip
\noindent
To prove Claim 2 we first consider the relative homology sequence for
the pair $V\supset \Sigma_0$,
$$
\cdots\quad\rightarrow H_1(\Sigma_0){\buildrel\alpha\over\rightarrow}
H_1(V){\buildrel\beta\over\rightarrow} H_1(V,\Sigma_0)
{\buildrel\partial\over\rightarrow}{\tilde H_0}(\Sigma_0)=0\quad .\eqno{(6)}
$$
(Here $\tilde H_0(\Sigma_0)$ is the reduced zeroeth-dimensional homology
group.) Since, as discussed in 
Lemma 4.4, $\alpha$ is onto,  we have
$\ker \beta ={\rm im}\alpha=H_1(V)$ which implies
$\beta
\equiv 0$. Hence $\ker\partial = {\rm im}\beta =0$, and thus $\partial$
is injective. This implies that $ H_1(V,\Sigma_0)=0$.

Now consider the relative homology sequence for the triple $V\supset
\partial V\supset \Sigma_0$,
$$
\cdots\quad\rightarrow H_1(\partial V,\Sigma_0) \rightarrow H_1 (V,\Sigma_0)
=0\rightarrow H_1(V,\partial V) {\buildrel\partial\over\rightarrow}
H_0(\partial V,\Sigma_0)\rightarrow \cdots\quad .\eqno{(7)}
$$
Since $H_0(\partial V,\Sigma_0)$ is torsion free and
$\partial$ is injective, $H_1(V,\partial V)$ is torsion free.
Next, Poincar\'e-Lefschetz duality gives $H^2(V)\cong H_1(V,
\partial V)$. Hence $H^2(V)$ is torsion free. The universal 
coefficient theorem implies that
$$H^2(V)\cong {\rm Hom}\, (H_2(V),Z)\oplus {\rm Ext}\, (H_1(V),Z)\quad .
\eqno{(8)}$$
The functor ${\rm Ext(-,-)}$ is bilinear in the first argument with 
respect to direct sums and ${\rm Ext(Z_k,Z)}=Z_k$. Hence $H^2(V)$ cannot 
be torsion free unless $H_1(V)$ is.  This completes the proof of
Claim 2 and the proof that  $H_*(V)$ is torsion free.

It remains to show that the inequalities in Lemma~4.3 become equalities.
We prove $b_2=k$; the equation $b_1 = \sum_{i=0}^k g_i$ then follows from 
equation (4). In view of Lemma~4.3, it is sufficient to show that $b_2 
\le k$. Since $H_2(V)$ is finitely generated and torsion free, we have 
$H_2(V)\cong H^2(V)\cong H_1(V,\partial V)$,
where we have again made use of Poincar\'e-Lefschetz duality.
Hence, $b_2 = {\rm rank}\, H_1(V,\partial V)$.  To show that
${\rm rank}\,H_1(V,\partial V)\le k$, we refer again to the long exact
sequence (7).
By excision,
$H_0(\partial V,\Sigma_0) \cong H_0(\partial V \setminus \Sigma_0,\emptyset)
=H_0(\partial V \setminus\Sigma_0)$.  Hence,
by the injectivity of $\partial$,  ${\rm rank}\,H_1(V,\partial V)$ $\le
{\rm rank}\, H_0
(\partial V,\Sigma_0)=$ the number of components of $\partial V \setminus
\Sigma_0 = k$.  This completes the proof of Proposition 4.5. \hfill\endproof
\smallskip

The conclusion of Proposition 4.5 applies to the spacelike
$3$-surface-with-boundary $V$ of Theorem 4.1.  Thus, we have completely
determined the
homology of the Cauchy surfaces of $\cal D_K'$.

We now consider the asymptotically flat case with null infinity $\scri =
\scri^+\cup \scri^-$.
For this case,
let $K$ be a spacelike cut of $\scri^-$, and let $\scri_K$ be the portion of
$\scri$
to the future of $K$, $\scri_K = \scri \cap J^+(K)$.  Let  $\cal D_K$ be
the domain of
outer of communications with respect to $\scri_K$,
 ${\cal D_K} =
I^+(\scri_K)\cap I^-(\scri_K)
= I^+(K) \cap I^- (\scri^+)$.
${\cal D}_K' := {\cal D}_K \cup \scri^+$ is a
connected spacetime-with-boundary, with $\partial \cal D_K' = \scri^+$
and $\cal D_K = \cal D_K' \setminus \scri^+$.

We then have the following analogue of Theorem 4.1.

\noindent {\bf Theorem 4.1$'$.} {\sl Let $\cal D_K$ be the domain of outer
communications to the future of the cut $K$ on $\scri^-$ of an asymptotically
flat spacetime $\cal M$ as described above.
Assume $\cal D_K'$ is  
globally hyperbolic and
satisfies the 
PTC. Suppose $V_0$ is a Cauchy surface for $\cal D_K$ such that its
closure $V = \overline{V_0}$ in $\cal M$
is a topological $3$-manifold-with-boundary, compact outside a small
neighborhood of $i^0$, with boundary components consisting  of
a disjoint union of compact $2$-surfaces,
$$\partial V =\bigsqcup\limits_{i=1}^{k}\Sigma_i\quad ,$$
where  the $\Sigma_i$,
$i=1,\dots,k$, are on the event horizon.
Then all $\Sigma_i$, $i= 1,\dots,k$ are $2$-spheres.  Moreover,
$V_0$ has the topology of a homotopy $3$-sphere minus $k+1$ closed $3$-balls.}

\smallskip
\noindent
{\bf Remark 1.\/} In the AF case the asymptotic topology is spherical, which
corresponds
to $g_0=0$ in inequality (3). But since $g_i=0$, $i=1,...,k$, inequality
(3) is satisfied in the AF
case,  as well.  Again, the topology
of the event horizon is constrained by the topology at infinity. 
\smallskip
\noindent
{\bf Remark 2.\/} This theorem
is a slightly strengthened version of the main theorem in [7]; it does
not assume orientability of $V_0$ and we conclude a stronger topology for this Cauchy
surface.
\smallskip
\noindent
{\bf Proof of Theorem 4.1$'$:}  The arguments used to prove
Theorem 4.1 can be easily adapted, with only minor technical changes
involved, to
prove in the AF case that each $\Sigma_i$ is a $2$-sphere.  Alternatively,
one may argue
as follows.  By known results on topological
censorship in the AF case ([5], [6]), $\cal D_K$ is simply
connected, and hence so is $V_0$.  It follows that $\tilde V = V \cup
\{i^0\}$ is a compact
simply connected
$3$-manifold-with-boundary, with boundary components $\S_i$, $i= 1,...,k$.
Then, according to Lemma 4.9, p.~47 in Hempel [23], each $\S_i$ is a 
$2$-sphere. By attaching $3$-cells to each $\S_i$ we obtain a closed simply 
connected $3$-manifold, which by well-known results (see [23]) is a homotopy
$3$-sphere.  Removing the attached $3$-cells and $i_0$ we obtain that
$V_0$ is a homotopy $3$-sphere minus $k+1$ closed balls.
\hfill\endproof

\smallskip
\noindent
{\bf Remark.\/}
Although the above results were proved 
assuming global hyperbolicity,
it is clear that the same results will hold
for a more general set of spacetimes that satisfy the PTC
and for which a version of Lemma 4.2 can be proved.
 Spacetimes that are not globally hyperbolic but
satisfy a weaker condition such as weak cosmic censorship can still admit 
a projection onto a preferred spacelike surface. In particular, one can 
generalize the projection given by the integral flow of a timelike vector 
field on the domain of outer communications used to push loops into the 
Cauchy surface to be a retract. Recall a retract of $X$ onto a subspace 
$A$ is a continuous map $r:X\to A$ such that $r|_A = \hbox{\rm id}$. Thus, 
if $V_0$ is a regular retract of $\cal D$, that is, if there exists a 
retract $ r:{\cal D} \cup {\scri} \to V_0 \cup \Sigma_0$ such that 
$r({\scri}) \subseteq\Sigma_0$, then one can again establish Lemma 4.2.

\vfil\eject
\noindent
{\bf V. Discussion}
\smallskip
\noindent
We wish to emphasize that the results concerning black hole topology
obtained in Section IV in no way contradict the numerical findings of
[24] concerning the existence in principle of temporarily toroidal
black holes in asymptotically flat spacetimes.
The consistency of topological censorship
with asymptotically flat models containing temporarily toroidal
black hole horizons has been clearly elucidated in [25].
The acausal nature of cross-over sets, expected to be present
in the early formation of the event horizon, permits
slicings of the event horizon in asymptotically flat
black hole spacetimes with exotic ({\it i.e.},
non-spherical) topologies.  See the recent
papers [26,27] for further discussion. As described in
Section IV, the method of topological censorship for
exploring the topology of black hole horizons makes use of
specific time slices, namely, Cauchy surfaces for the DOC 
or for the sub-region of the DOC to the future of a cut on 
$\scri$.
Surfaces  exhibiting temporarily toroidal black hole horizons are not 
such surfaces.
Moreover, the method requires such a slice to have non-empty edge which
meets the horizons in $C^0$ compact surfaces.
We elaborate further on these points below.

It is important to keep in mind that not all Cauchy surfaces $V_0$ for
the DOC are
interiors of orientable manifolds with boundary $V$ corresponding to
the intersection of a spacelike slice with the black hole horizons. 
Consider 
the $t=0$ slice of the $RP^3$ geon. As discussed in [5],
this spacetime is constructed from the $t=0$ slice of Schwarzschild 
spacetime by identifying antipodal points at the throat $r=2M$. The 
maximal evolution of this slice is a spacetime with spatial topology 
$RP^3 -pt$. Its universal covering space is the maximally extended 
Schwarzschild spacetime.  

The $t=0$ slice of the $RP^3$ geon contains a non-orientable $RP^2$ 
with zero expansion. This $RP^2$ is not a trapped surface
as it does not separate the slice into two regions. It is not part of 
the DOC as any radially outward directed null 
geodesic from this surface does not intersect $\scri$; thus it is clearly 
part of the horizon. The intersection of the DOC
with the $t=0$ slice produces a simply connected $V_0$. The intersection 
of the horizon with this slice is $RP^2$. However, we cannot attach this 
surface to $V_0$ to produce a manifold with boundary $V$ by the inclusion 
map; instead, this map reproduces the original $t=0$ slice which has no 
interior boundary. Note, however, that any spacelike slice that does
not pass through this $RP^2$ will intersect the horizon at an $S^2$. 
In fact, this will be the generic situation. Moreover, the intersection 
of such a slice with the DOC will produce a 
simply connected $V_0$ which is the interior of a closed connected 
orientable $V$ with an $S^2$ interior boundary.
 
Clearly this example does not contradict any results of Section IV, which 
assumes an orientable $V$ with two or more boundaries. However, it does 
yield the important lesson that one must construct $V$ to apply the theorem, 
not $V_0$. It also gives an example of a badly behaved cut of the horizon, 
again illustrating the usefulness of taking slices to the future of a cut of 
$\scri$.

For our second example, we construct a toy model of a black 
hole spacetime that mimics a special case of topology change, namely that 
of black hole formation from a single collapse. This model illustrates
several features; how a choice of a hypersurface 
can affect the description of horizon topology and how some cuts of
$\scri$ give rise to a Cauchy surface whose edge on the horizon is not a $2$-manifold.

We begin with a 3-dimensional model; later a 4-dimensional example 
will be constructed by treating the 3-dimensional model as a 
hyperplane through an axis of symmetry in the larger spacetime.
Our spacetime can have either anti-de Sitter or flat geometry; 
as both are conformal to regions of the Einstein static cylinder,
we use as coordinates in the construction below
those of the conformally related flat metric
( {\it cf.} [19], sections 5.1 and 5.2). 
We depict $\scri$ as timelike in the accompanying 
figures, but it can equally well be null.

We begin in 3 dimensions with a line segment $L$ defined by $t=y=0$, $|x|\le l$.
The future $I^+(L)$ of this line segment is a sort of elongated cone, whose traces 
in hyperplanes of constant $t$ have the shape of rectangles with semi-circular 
caps attached to the two short sides. We foliate the spacetime by hyperboloids, 
$$t=a+\sqrt{r^2+b}\quad ,\eqno{(9)}$$
where $r^2=x^2+y^2$ and $a,b$ are conveniently chosen parameters. 
These hyperboloids cut $\scri$ in a circle.
We now remove from 
spacetime all points of 
$J^+(L)$ above a 
hyperplane 
$t = d$ intersecting 
it to the future. What remains is a black hole spacetime and has a globally 
hyperbolic domain of outer communications $\cal D$ (again, {\it cf.} Section II). 
The black hole is the 
set of remaining points of $I^+(L)$. The horizon $\partial I^-({\scri}^+)$ 
is generated by null geodesics that all begin on $L$. 

The Cauchy surface for $\cal D$ will have topology $R^2$ and does not cross 
the horizon. Thus, to probe the topology of the horizon, one needs to consider 
spacetimes corresponding to the future of a cut of $\scri$. However, not every 
cut of $\scri$ will produce a spacetime with a Cauchy slicing with the correct 
properties. Such a bad cut of $\scri$ is illustrated in Figure 1. 
The boundary of the causal future of this cut 
intersects the horizon at a segment $I$ of $L$. The topology of a 
Cauchy slice for its DOC is $R^2 \setminus I$. Its closure intersects the 
horizon at $I$; thus, as in the $RP^3$ geon above, 
the closure of this slice has no inner boundary, being in this case $R^2$.

This 3-d spacetime corresponds to a 4-d axisymmetric spacetime. 
The correspondence between axisymmetric spatial hypersurfaces and the
$xy$ planes of the spacetime is generated by rotating each
$xy$ plane about the $y$ axis. 
After this rotation, one sees that
 the line segment $L$ becomes a disk in the 4-d axisymmetric spacetime.
The Cauchy surface in question 
meets the horizon in a disk (a closed 2-ball) and has topology $R^3$ minus 
that disk. The closure of the Cauchy surface is $R^3$ and again has 
no internal boundary. Thus the results of Section IV do not apply to this 
Cauchy slice.

A good cut of $\scri$ is illustrated in Figure 2. This cut 
intersects the horizon at a sphere. The topology of a Cauchy surface for its 
DOC is $R^2\setminus {\overline {B^2}}$ and the intersection of its closure is 
$S^1$. 
The Cauchy surface in the corresponding 4-dimensional model has
topology  $R^3\setminus 
{\overline {B^3}}$ 
with internal boundary $S^2$. The results of Section 
IV clearly apply in the latter case.

Of course, not all spacelike surfaces need be Cauchy surfaces of the spacetime
to the future of a cut of $\scri$. The family of hyperboloids (9), two of which are
illustrated in Figure 3,
 provides an  example of such surfaces.
As recognized in [25] in a similar 
model, this family exhibits formation of a temporarily toroidal black hole 
horizon as the parameter $a$ in (9) increases; these surfaces intersect the 
horizon in a pair 
of topological circles, which by axial symmetry correspond to a toroidal 
horizon in the 4-dimensional spacetime. The circles increase in size and 
eventually meet, whence the horizon topology changes. After this point, 
these surfaces meet the horizon at a circle, corresponding to 
a sphere by axial symmetry.

In contrast, with respect to constant-$t$ surfaces, the 
horizon forms completely at the $t=0$ instant.
For every $t>0$ hypersurface, the black hole has spherical topology,
and inequality (3) holds. Any $t=t_0>0$ hypersurface is a Cauchy surface
for the region of $\cal D$ that lies in the future of an appropriate cut of
$\scri$. 
The apparent change of horizon topology from toroidal to spherical was 
an effect entirely dependent on the choice of hypersurface. The
only unambiguous description of this black hole is that no causal curve
was able to link with the horizon; {\it i.e.}, that the PTC was not 
violated.

In the introduction, we offered the view that the topology of
the boundary at infinity constrained that of the horizons, but one
could equally well reverse this picture. Let us contemplate a black
hole considered as a stationary, causally well-behaved, isolated system
cut off from the Universe by a sufficiently distant boundary---$\scri$.
Then we have shown that topological censorship requires the genus of
the horizon to be a lower bound for that of the boundary. As remarked
above, 
this seems an intuitive result; 
as illustrated in Figure 4,
when one visualizes placing a genus $g_1$ surface within a genus $g_2$ 
``box'' with no possibility of
entangling curves, it seems clear that $g_2\ge g_1$. Yet, as is often
the case with such things, the powerful machinery of algebraic topology
was required to prove it. An advantage of using this powerful tool is
that we were able to completely specify the homology of
well-behaved exterior regions of black holes and, in virtue of
Proposition 4.5, to say that all interesting homology of these
exteriors, save that which is reflected in the topology of scri, is
directly attributable to the presence of horizons.

\vfil\eject
\noindent
{\bf Acknowledgements}
\smallskip
\noindent
EW wishes to acknowledge conversations with R.B.~Mann and W.~Smith
concerning locally anti-de Sitter spacetimes. GJG and EW thank Piotr
Chru\'sciel for conversations concerning the validity of topological
censorship for these spacetimes. This work was partially supported by
the Natural Sciences and Engineering Research Council of Canada and
by the National Science Foundation (USA), Grant No.~DMS-9803566.

\bigskip
\noindent
{\bf References}
\smallskip
\noindent
\item{[1]}{S.W.~Hawking, {\it Commun.~Math.~Phys.}~{\bf 25} (1972), 
152.}
\item{[2]}{G.J.~Galloway, in {\it Differential Geometry and Mathematical
Physics}, {\it Contemporary~Math.} {\bf 170}, ed.~J.~Beem and K.~Duggal
(AMS, Providence, 1994), 113.}
\item{[3]}{P.T.~Chru\'sciel and R.M.~Wald, {\it Class.~Quantum 
Gravit.}~{\bf 11} (1994), L147.}
\item{[4]}{S.F.~Browdy and G.J.~Galloway, {\it J.~Math.~Phys.}~{\bf 36} 
(1995), 4952.}
\item{[5]}{J.L.~Friedman, K.~Schleich, and D.M.~Witt, 
{\it Phys.~Rev.~Lett.}~{\bf 71} (1993), 1486.}
\item{[6]}{G.J.~Galloway, {\it Class.~Quantum Gravit.}~{\bf 12} (1995),
L99.}
\item{[7]}{T.~Jacobson and S.~Venkataramani, {\it Class.~Quantum
Gravit.}~{\bf 12} (1995), 1055.}
\item{[8]}{J.P.S.~Lemos, {\it Phys.~Lett.}~B {\bf 352} (1995), 46.}
\item{[9]}{M.~Ba\~nados, C.~Teitelboim, and J.~Zanelli, 
{\it Phys.~Rev.~Lett.}~{\bf 69} (1992), 1849; M.~Ba\~nados, 
M.~Henneaux, C.~Teitelboim, and J.~Zanelli, {\it Phys.~Rev.}~D{\bf 48}
(1993), 1506.}
\item{[10]}{S.~\AA minneborg, I.~Bengtsson, S.~Holst, and P.~Peld\'an,
{\it Class.~Quantum Gravit.}~{\bf 13} (1996), 2707.}
\item{[11]}{R.B.~Mann, {\it Class.~Quantum Gravit.}~{\bf 14} (1997),
2927}
\item{12}{W.L.~Smith and R.B. Mann, {\it Phys.~Rev.}~D {\bf 56} (1997),
4942.}
\item{[13]}{D.R.~Brill, J.~Louko, and P.~Peld\'an, 
{\it Phys.~Rev.}~D {\bf 56} (1997), 3600.}
\item{[14]}{R.B. Mann, in {\it Internal Structure of Black Holes and 
Spacetime Singularities}, eds.~L. Burko and A.~Ori, 
{\it Ann.~Israeli Phys.~Soc.}~{\bf 13} (1998), 311.}
\item{[15]}{G.J.~Galloway, {\it Class.~Quantum Gravit.}~{\bf 13} (1996),
1471.}
\item{[16]}{G.J.~Galloway and E.~Woolgar, {\it Class.~Quantum 
Gravit.}~{\bf 14} (1997), L1.}
\item{[17]}{S.~\AA minneborg, I.~Bengtsson, D.~Brill, S.~Holst, and 
P.~Peld\'an, {\it Class.~Quantum Gravit.}~{\bf 15} (1998), 627.}
\item{[18]}{L. Vanzo, {\it Phys.~Rev.}~D {\bf 56} (1997), 6475.}
\item{[19]}{S.W.~Hawking and G.F.R.~Ellis, {\it The Large Scale 
Structure of Space-Time}, (Cambridge University Press, Cambridge, 1973).}
\item{[20]}{E.~Woolgar, in progress.}
\item{[21]}{R.D.~Sorkin and E.~Woolgar, 
{\it Class.~Quantum Gravit.}~{\bf 13} (1996), 1971.}
\item{[22]}{B.C.~Haggman, G.W.~Horndeski, and G.~Mess, 
{\it J.~Math.~Phys.}~{\bf 21} (1980), 2412.}
\item{[23]}{J.~Hempel, {\it 3-Manifolds}, (Princeton University Press, 
Princeton, 1976).}
\item{[24]}{S.A.~Hughes et al., {\it Phys. Rev.}~D {\bf 49} (1994), 4004.}
\item{[25]}{S.L.~Shapiro, S.A.~Teukolsky, and J.~Winicour, {\it
Phys.~Rev.}~D {\bf 52} (1995), 6982.}
\item{[26]}{S.~Husa and J.~Winicour, preprint gr-qc/9905039.}
\item{[27]}{M.~Siino, {\it Phys.~Rev.}~D {\bf 58} (1998), 104016;
D {\bf 59} (1999), 064006.}
\vfill\eject
\noindent
{\bf Figures}
\smallskip
\input epsf
\hsize=160 true mm
\vsize=250 true mm

\nopagenumbers
\midinsert
\vskip .1in
\hskip 0in
\epsfxsize=4.8in
\epsfysize=4.8in
\epsfbox{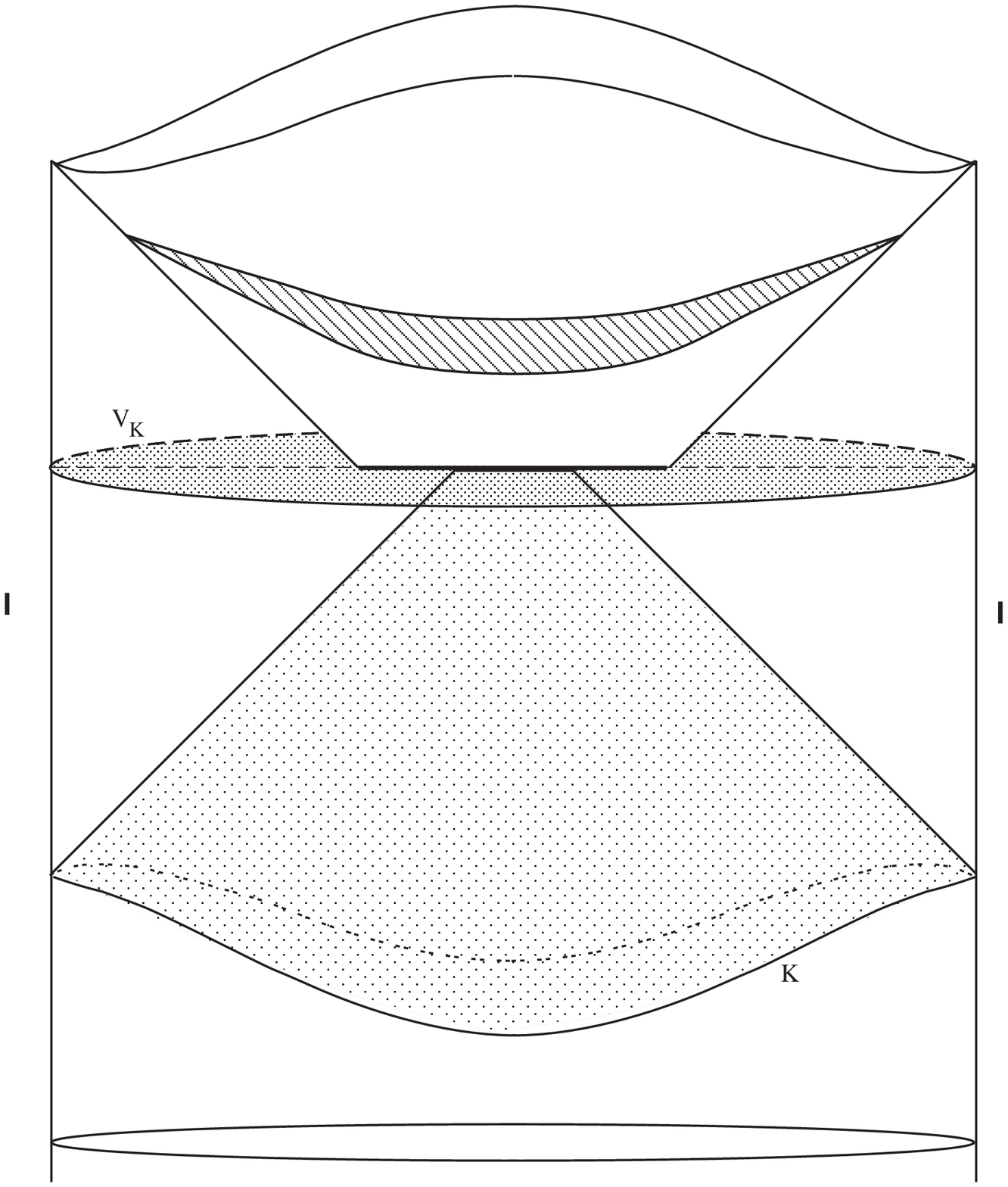}
\endinsert
\vskip 0.2in
\noindent
Figure 1. A bad cut of the horizon. The surface $V_K$ is 
the Cauchy surface for the domain of outer communications of the spacetime
to the future of the cut $K$ of $\scri$. The inner boundary of the causal
future of $K$ intersects the horizon at a spacelike line segment.
 Consequently the closure of $V_K$ has topology $R^2$.
\vfill \eject 

\midinsert
\vskip 1in
\hskip 0in
\epsfxsize=4.8in
\epsfysize=4.8in
\epsfbox{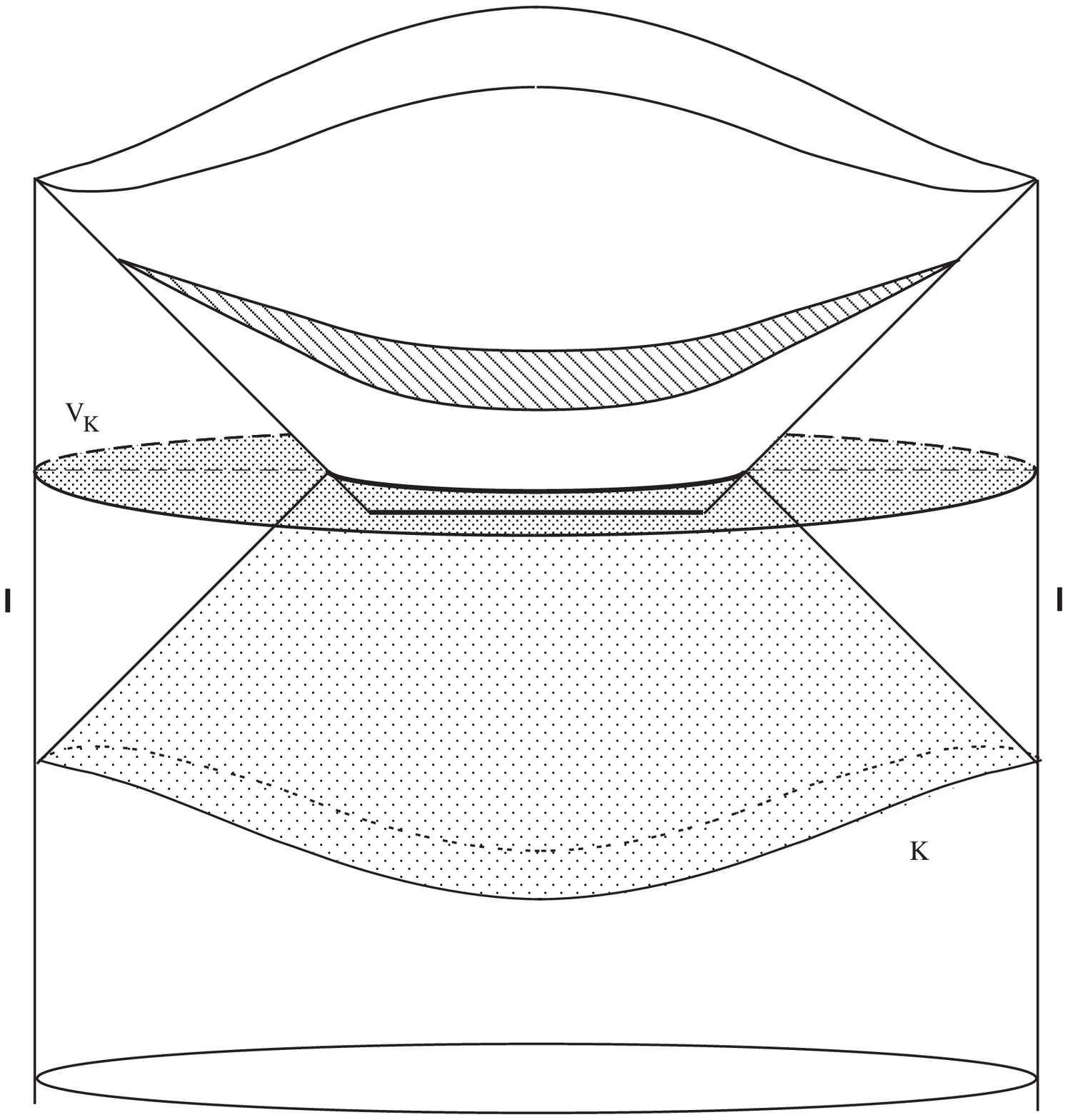}
\endinsert
\vskip .5in
\noindent
Figure 2. A good  cut of the horizon. The surface $V_K$ is
the Cauchy surface for the domain of outer communications of the spacetime
to the future of the cut $K$ of $\scri$. The intersection of
the inner boundary of the causal
future of  $K$ with the horizon is now $S^1$. Consequently the closure
of $V_K$ intersects the horizon at $S^1$.
\vfill \eject 

\midinsert
\vskip 0in
\hskip 0in
\epsfxsize=4.8in
\epsfysize=4.8in
\epsfbox{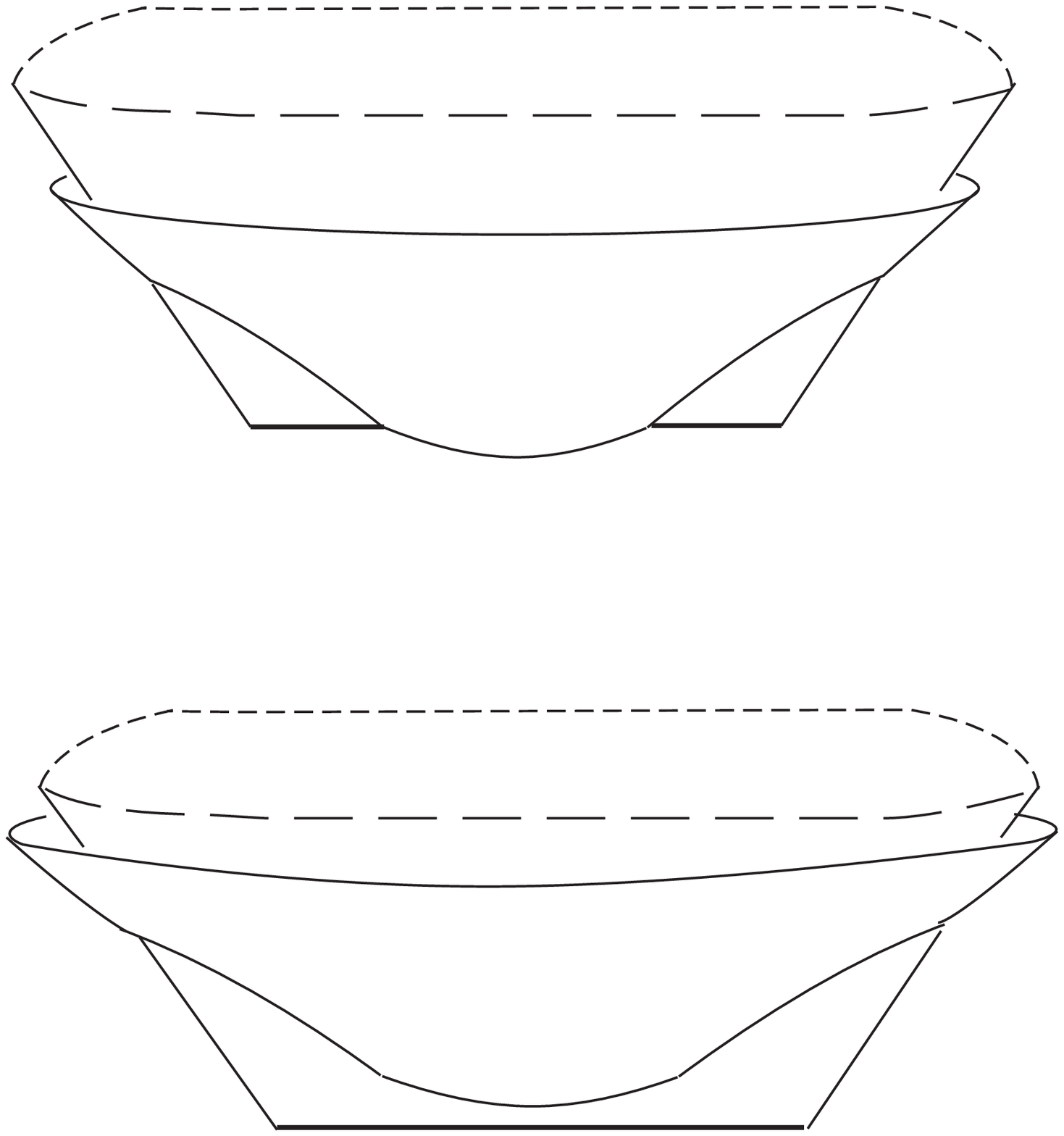}
\endinsert
\vskip 1.0in
\noindent
Figure 3. Two slicings of the horizon by hyperboloids. Both illustrations concentrate
on the region near the horizon. The top illustration
is of a hyperboloid that intersects the horizon at two topological circles.
The bottom illustration is of a hyperboloid that lies to the future of the first.
It intersects the horizon at one topological circle.

\vfill \eject 
\midinsert
\vskip 0in
\hskip 0.8in
\epsfxsize=2.75in
\epsfysize=5.5in
\epsfbox{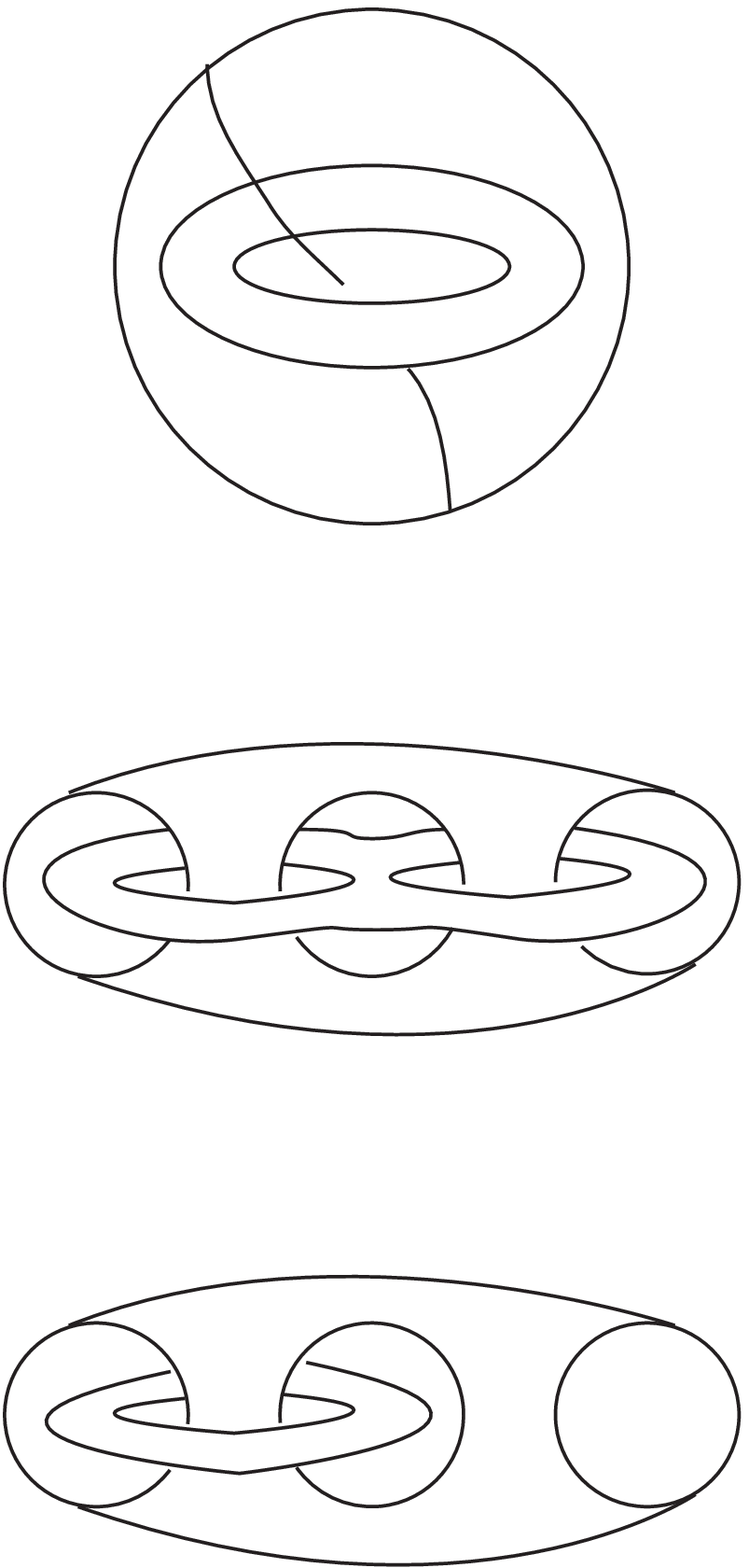}
\endinsert
\vskip 0.2in
\noindent
Figure 4. Entanglement and non-entanglement of curves
on black hole horizons.
Each illustration displays a cross-section of a cut of $\scri$. 
The cut-away reveals a cut of a black hole horizon inside.
In the top illustration, the cut of $\scri$ has genus zero, that of
the horizon has genus 1, and, as illustrated,
there are curves not deformable to $\scri$.
In the middle illustration, the genus of the cut of $\scri$ and that of the horizon are both
2, and they are linked in such a manner that every curve is deformable to
$\scri$. In the illustration at bottom, the genus of the cut of $\scri$ exceeds that of the
horizon, and again every curve is deformable to $\scri$.
\vfill\eject
\bye